\documentclass[aps,prl,twocolumn,showpacs,superscriptaddress,preprintnumbers]{revtex4-2}

\usepackage{style}
\usepackage{calrsfs}
\DeclareMathAlphabet{\pazocal}{OMS}{zplm}{m}{n}
\newcommand{\La}{\mathcal{L}}

\begin{document}

\preprint{FERMILAB-PUB-23-539-T}
\preprint{MIT-CTP/5649}

\title{New $\mu$ Forces From  $\nu_\mu$ Sources}

\author{Cari Cesarotti \,\orcidlink{0000-0001-5128-7919}}
\email{ccesar@mit.edu}
\affiliation{
Center for Theoretical Physics, Massachusetts Institute of Technology,
Cambridge, MA 02139, USA}

\author{Yonatan Kahn \,\orcidlink{0000-0002-9379-1838}}
\email{yfkahn@illinois.edu}
\affiliation{Illinois Center for Advanced Study of the Universe and Department of Physics, University of Illinois Urbana-Champaign, Urbana, IL 61801}

\author{Gordan Krnjaic\,\orcidlink{0000-0001-7420-9577}} 
\email{krnjaicg@fnal.gov}
\affiliation{Theoretical Physics Department, Fermi National Accelerator Laboratory, Batavia, Illinois 60510}
\affiliation{Department of Astronomy and Astrophysics, University of Chicago, Chicago, IL 60637}
\affiliation{Kavli Institute for Cosmological Physics, University of Chicago, Chicago, IL 60637}

\author{Duncan Rocha \,\orcidlink{0000-0001-7420-9577}} 
\email{krnjaicg@fnal.gov}
\affiliation{Kavli Institute for Cosmological Physics, University of Chicago, Chicago, IL 60637}

\author{Joshua Spitz \,\orcidlink{0000-0002-6288-7028}}
\email{spitzj@umich.edu}
\affiliation{Department of Physics, University of Michigan, Ann Arbor, MI 48109}

\date{\today}

\begin{abstract}
 Accelerator-based experiments reliant on charged pion and kaon decays to produce muon-neutrino beams also deliver an associated powerful flux of muons. Therefore, these experiments can additionally be sensitive to light new particles that preferentially couple to muons and decay to visible final states on macroscopic length scales. Such particles are produced through rare 3-body meson decays in the decay pipe or via muon scattering in the beam dump, and decay in a downstream detector. 
 To demonstrate the potential of this search strategy, we recast existing MiniBooNE and MicroBooNE studies of neutral pion production in neutrino-induced neutral-current scattering ($\nu_\mu N \to \nu_\mu N \pi^0,~\pi^0\rightarrow \gamma\gamma$) to place new leading limits on light ($< 2m_\mu$) muon-philic scalar particles that decay to diphotons through loops of virtual muons. Our results exclude scalars of mass between 10 and 60 MeV  in which this scenario resolves the muon $g-2$ anomaly. We also make projections for the sensitivity of SBND to these models and provide a road map for future neutrino experiments to perform dedicated searches for muon-philic forces.

\end{abstract}

\maketitle

\section{Introduction}
\label{sec:intro}

Light, weakly-coupled particles arise in many well-motivated extensions of the Standard Model (SM) and may shed light on several open questions in particle physics, including the particle nature of dark matter \cite{Berlin:2018bsc}, the hierarchy problem \cite{Morrissey:2009ur}, and the strong CP problem \cite{Dine:1981rt}.
Motivated by these fundamental questions, there is a global program of accelerator searches to discover these new particles (see Ref.~\cite{Gori:2022vri} for a review).
However, most searches involve beams of protons or electrons, which are naively limited in their sensitivity to particles that couple preferentially to heavy flavor. 

\begin{figure*}[t]
    \includegraphics[width=5.2 in]{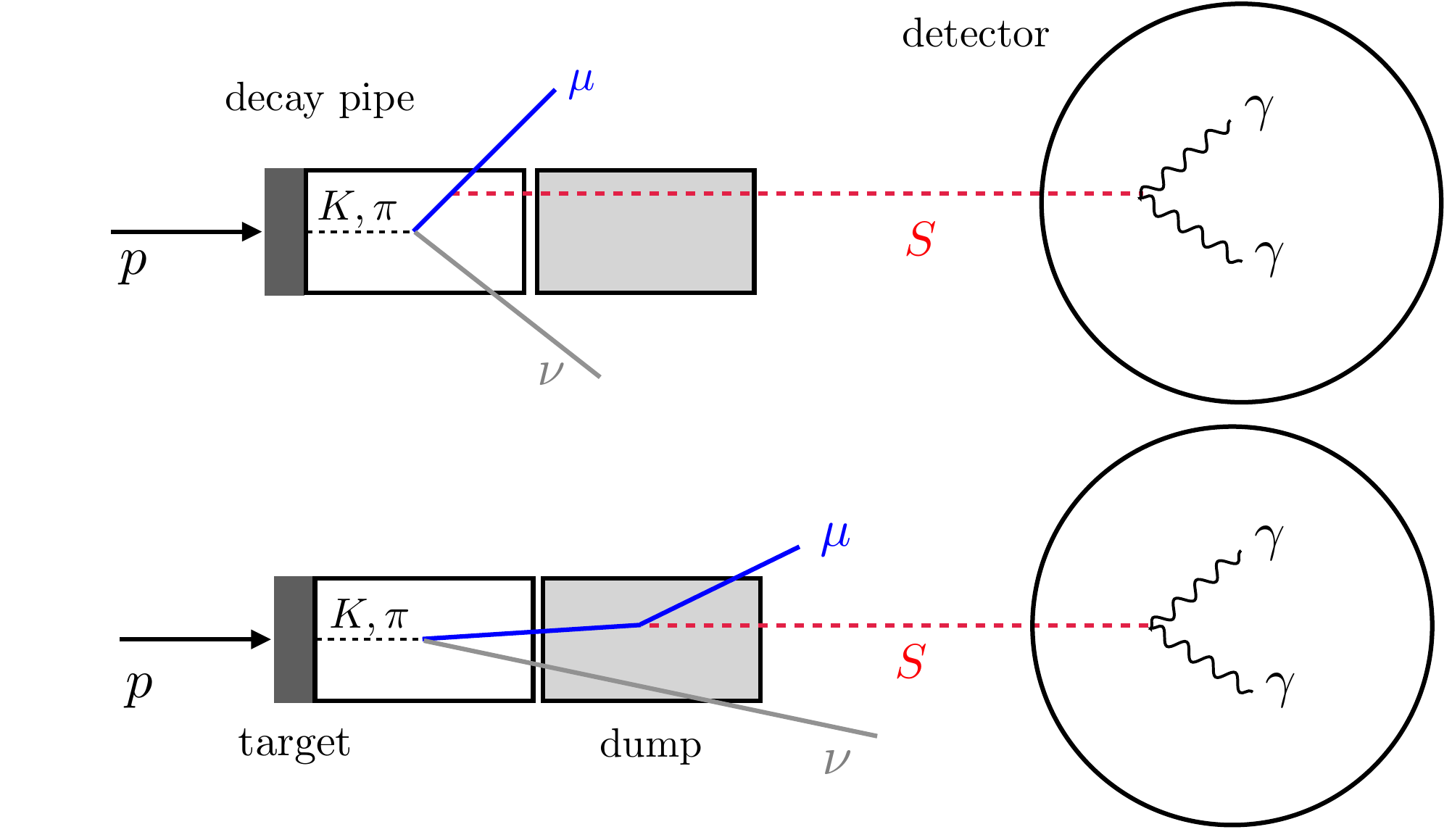}
    \caption{Schematic cartoon of our MiniBooNE and MicroBooNE signal processes. {\bf Top:} the 8 GeV proton beam impinges on a fixed Be target and produces mesons including charged kaons, whose decays yield a flux of muons and associated scalars $S$ through $\pi^\pm, K^\pm\to \mu^\pm \nu_\mu S$ decays. {\bf Bottom:} $S$ may also be created via a spray of pions/kaons whose decays yield a secondary muon beam in the decay pipe. The muons then interact with the beam dump, where they produce long-lived muon-philic scalar particles $S$ through bremsstrahlung-like processes. Note that the kaon production mode is independent of the beam dump, while the muon bremsstrahlung mode requires scattering in the beam dump. In both cases, for $m_S < 2m_\mu$, the scalars eventually decay to diphotons in the downstream detector. For details of the experimental setup and detector performance, see Ref.~\cite{MiniBooNE:2008paa}.}
    \label{fig:cartoon}
\end{figure*}

Over the past decade there have been several proposals for dedicated muon-beam experiments.
Studying muons directly is of interest as many experimental milestones remain unexplored or unexplained, such as muon-philic forces \cite{Bauer:2018onh,Capdevilla:2021kcf,Ilten:2018crw}, dark matter \cite{Berlin:2018bsc,Kahn:2018cqs,Holst:2021lzm}, and the persistent $g-2$ anomaly \cite{Muong-2:2023cdq}.
If the new muon-philic particles are invisible on accelerator length scales, the proposed NA64$\mu$ \cite{Gninenko:2014pea,Chen:2018vkr} and $M^3$ \cite{Kahn:2018cqs} fixed-target experiments can leverage missing energy and momentum to probe key targets related to these questions. 
 Conversely, if the new states decay visibly as long-lived particles on
 accelerator scales, they can be probed with new beam-dump experiments \cite{Chen:2017awl,Cesarotti:2022ttv,Cesarotti:2023sje}.
 It has also been shown that a future muon collider program
 can comprehensively expand the reach for a variety of decay channels and event topologies, particularly for heavier states \cite{MuonCollider:2022nsa}.

However, we do not have to wait for a dedicated muon beam facility to study such interactions.
Since relativistic proton scattering can efficiently produce mesons which decay to secondary muons, existing data can be used to probe muon-philic forces. These muons can source new particle production directly 
through radiative  meson decays or through secondary muon interactions 
in macroscopic material.
%
%
For example, kaon decays at NA62 can source new forces as final-state radiation from muon lines \cite{Krnjaic:2019rsv}, and the muons produced through the Drell-Yan process at ATLAS can exhibit kinked 
tracks if they emit new invisible particles by scattering off detector constituents \cite{Galon:2019owl}. 
Alternatively, secondary muons produced in fixed-target proton collisions at SpinQuest/DarkQuest can interact with the target to yield new states that decay visibly and can be constrained using the downstream tracker \cite{Forbes:2022bvo}.

In this {\it Letter}, we show that {\it neutrino} sources are powerful probes of new muon-philic particles. As a proof of concept, we place new limits on long-lived  muon-philic particles using existing MiniBooNE and MicroBooNE searches for neutral pion production in neutral current neutrino scattering, $\nu N \to \nu N 
\pi^0$ \cite{MiniBooNE:2009dxl,MicroBooNE:2022zhr}. Based on a conservative analysis of these data, we improve existing
constraints on light scalars that couple exclusively to the muon and decay to diphotons at 
loop level.
Notably, as shown in Fig.~\ref{fig:results}, our limits already exclude almost an order of magnitude in scalar mass over which this scenario could resolve
the muon $g-2$ anomaly, parameter space which was previously targeted by a proposed dedicated muon beam-dump experiment~\cite{Chen:2017awl}. 

\vspace{-0.45cm}

\section{Model Overview}
\vspace{-0.25cm}
\label{sec:theory}
We consider a muon-philic scalar $S$, a SM singlet of mass $m_S$ with the Yukawa interaction 
\vspace{-0.05cm}
\be
\label{eq:lag}
\vspace{-0.25cm}
\La_{\rm int} \supset y S \bar \mu \mu~,
\vspace{-0.2cm}
\ee
which we have written in four-component Dirac fermion notation. 
This interaction can arise from a 
gauge-invariant dimension-5 operator 
\be
\La_{\rm eff} \supset \brac{y}{v} S H^\dagger L_2 \mu^c + {\rm h.c.}
\ee
where $H$ is the SM Higgs doublet, $v = 246$ GeV is its vacuum expectation value,  $L_2$ is the second generation lepton $SU(2)_L$ doublet, and $\mu^c$ is the $SU(2)_L$ singlet muon field. This non-renormalizable interaction may be UV-completed with
a heavy vector-like fourth generation of leptons, which mixes with the muon and is integrated out in the low-energy effective theory \cite{Batell:2017kty,Batell:2021xsi,Egana-Ugrinovic:2019wzj}.

\begin{figure*}[t]
    \includegraphics[width=5.5 in]{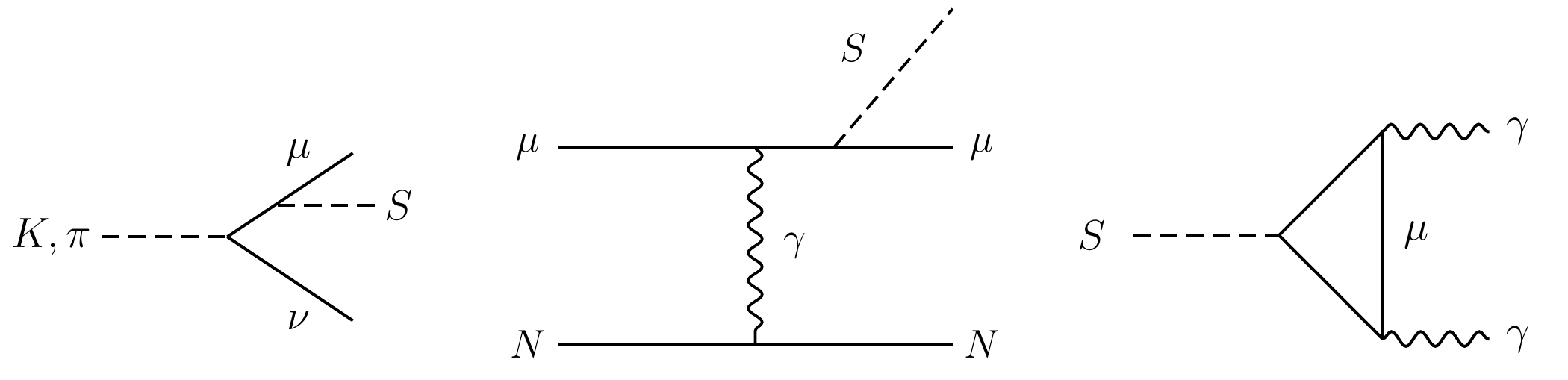}
    \vspace{0.51cm}
    \caption{ 
    {\bf Left:} Feynman diagram for scalar production from 
    $K,\pi \to \mu \nu S$ decays. 
    {\bf Middle:} Scalar production via
    $\mu$-$N$ scattering in 
    the beam dump. {\bf Right:} Scalar decay through a muon loop using only
     the coupling in \Eq{eq:lag}.}
    \label{fig:Feynman}
\end{figure*}

\begin{figure*}[t]
    \includegraphics[width=7 in]{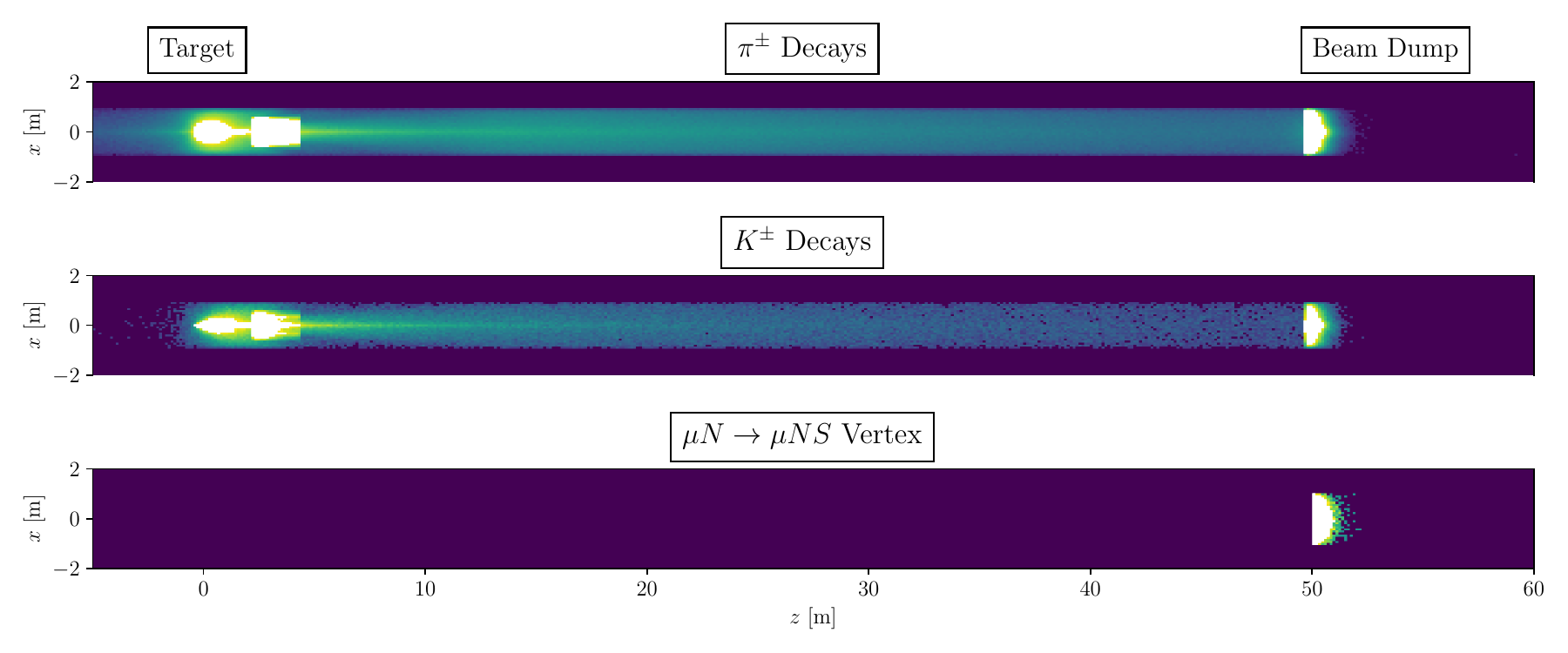}
    \caption{Heatmaps (arbitrary units) of the processes in the beam dump that produce muon-philic scalars, for $r_\perp < 0.9$\,m, where $r_\perp$ is the cylindrical radius in the $x-y$ plane. This displays the scalar production inside the decay pipe, however the analysis considers production events that occur outside the pipe as well. The top (middle) diagram shows the vertices where the $\pi^{\pm}$ ($K^{\pm}$) decays, which may contain scalars in their decay products. The lower diagram shows the spatial distribution of scalar emission from muons scattering in the beam dump, specifically for $m_S = 100$\,MeV, although this distribution does not vary significantly over the scalar mass. The plot has several features: the target can be seen on the far left, with the electromagnet adjacent, and the beam dump is visible on the right. }
    \label{fig:geant}
\end{figure*}

If the operator in \Eq{eq:lag} is the scalar's only coupling to SM fermions, then for $m_S < 2m_\mu$, $S$ will decay to diphotons through muon loops. The width for this process is~\cite{Chen:2017awl}
\be\label{eq:width}
\Gamma_{S} = 
\frac{\alpha^2 y^2 m_S^3}{64 \pi^3 m_\mu^2} x^2 \left|  1 + (1-x) f(x)   \right|^2, 
\ee
where $x \equiv 4 m_\mu^2/m_S^2$ and  $f(x) = \arcsin^2(x^{-1/2})$ for $x > 1$. Note that for nearly all viable parameter values in this low mass range, $S$ is a long-lived particle with a decay length
\be
\frac{ c\tau_S}{\gamma} \approx 300 \,{\rm m} \brac{10^{-4}}{y}^2
\brac{40 \, \rm MeV}{m_S}^3,
\label{eq:approxD}
\ee
where $\tau_S = 1/\Gamma_{S}$ is the $S$ lifetime and $\gamma$ is the boost factor. This estimate uses \Eq{eq:width} in the $x \gg 1$ limit. 

In our parameter space of interest (see Fig.~\ref{fig:results}), the Yukawa coupling in \Eq{eq:lag} always satisfies $y \gtrsim 10^{-5}$, which is
sufficient to bring $S$ into chemical equilibrium with SM particles in the early universe. Therefore, in order to avoid cosmological limits on extra light degrees of freedom during Big Bang nucleosynthesis, we generically require $m_S \gtrsim$ few MeV \cite{Capdevilla:2021kcf,Escudero:2019gzq}.

\section{Analysis}
Our search concept leverages secondary muon production
at MiniBooNE and MicroBooNE following a sequence of steps,
depicted schematically in Fig.~\ref{fig:cartoon} with relevant Feynman diagrams in Fig.~\ref{fig:Feynman}: 

\begin{enumerate}
    \item The 8 GeV proton beam from the Fermilab Booster Neutrino Beamline (BNB) strikes the Be target, producing a forward shower of pions and kaons;
    \item The charged pions and kaons are focused down the beamline via an electromagnet, called the \textit{horn}, and then typically decay in flight in the decay pipe;
    
    \item New scalar singlets $S$ can be produced either directly 
    through radiative 3-body  $\pi^{\pm}, K^{\pm} \to \mu^{\pm} \nu_\mu S$ decays (Fig. \ref{fig:cartoon}, top), or through $\mu N \to \mu N S$ scattering  from the secondary interactions in the beam dump (Fig. \ref{fig:cartoon}, bottom);
    
    \item The $S$ particles propagate on-shell and decay visibly to $\gamma \gamma$  in the downstream detector.
\end{enumerate}

To perform our numerical analysis, we utilize samples of pions, kaons, and muons from $10^7$ protons-on-target (POT) created from a custom \texttt{GEANT4}~\cite{Geant4} simulation of the BNB in neutrino mode. The detailed simulation includes all relevant beamline elements -- the target, horn electromagnet and magnetic field, collimator, decay pipe, shielding, and beam dump -- and is based on the description in Refs.~\cite{MiniBooNE:2008paa,MiniBooNE:2008hfu}. The $10^7$ POT produce $\sim 10^6$ charged kaons and $\sim 2 \times 10^7$ muons (primarily via decays of pions produced at the target).

\begin{figure}[t]
\vspace{-0.35cm}
    \includegraphics[width=3.5 in]{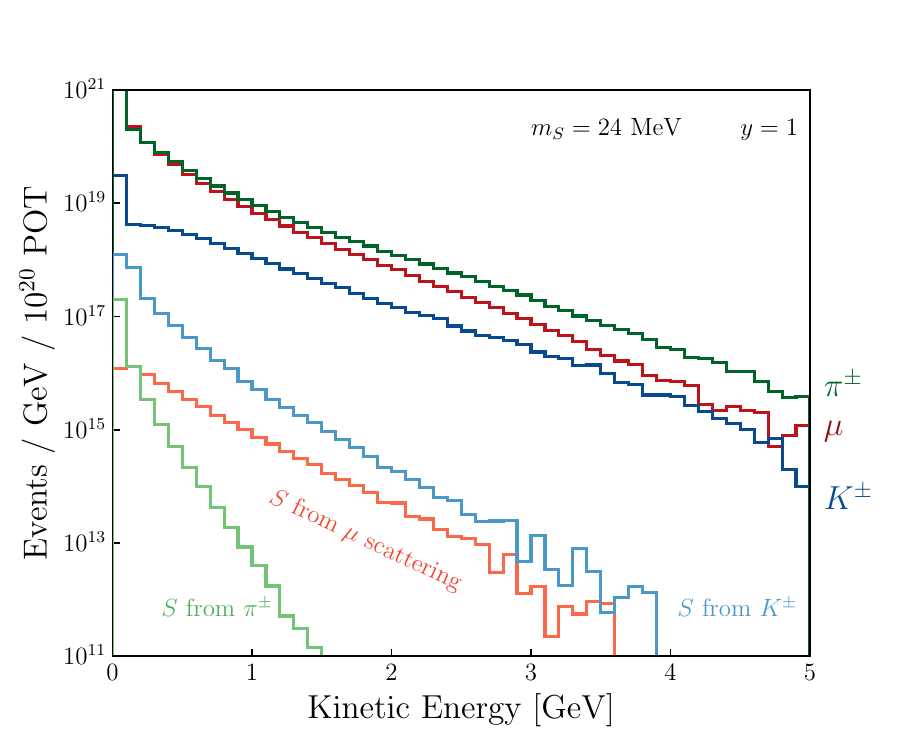}
    \caption{Energy distribution of charged pions (dark green), charged kaons (dark blue), and secondary 
    muons (dark red) produced from proton-Be collisions at 8 GeV. Also shown are energy distributions for scalars produced from pion decays (light green), kaon decays (light blue) and from  secondary muon scattering in the beam dump (orange). The scalar distributions are normalized to $y=1$ to enhance visibility on the plot, but in our analysis we explore the region $y \in [10^{-5}, 10^{-2}]$. These curves were
    generated with a \texttt{GEANT} simulation
    utilizing the setup described in Refs.~\cite{MiniBooNE:2008paa,MiniBooNE:2008hfu}. 
    } 
    \label{fig:distribution}
\end{figure}
\begin{figure}[t!]
\vspace{-0.5cm}
\hspace{-0.5cm}
    \includegraphics[width=3.5 in]{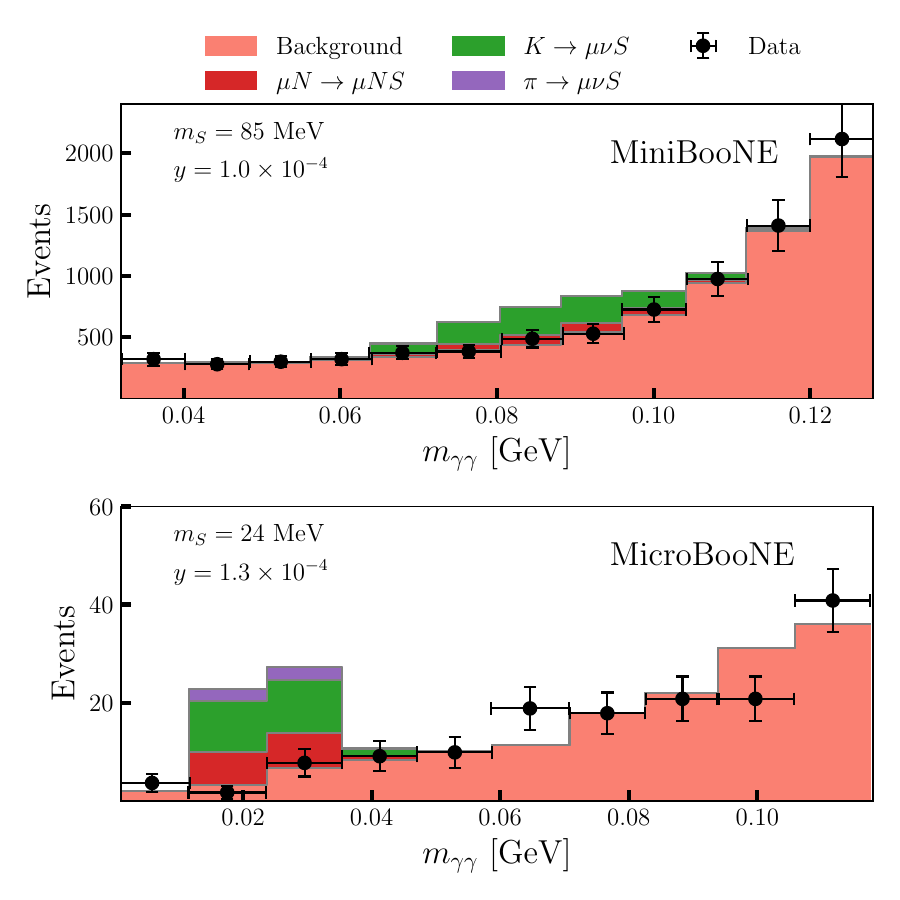}~
    \caption{Example signal distributions generated by scalar decays in the MiniBooNE and MicroBooNE detectors. The data points are measured $\gamma \gamma$-like events, and the background prediction arises primarily from neutrino-induced $\pi^0\to \gamma \gamma$ events, and digitized from \cite{MiniBooNE:2009dxl, MicroBooNE:2022zhr}. The MiniBooNE error bars include statistical and systematic uncertainties. The MicroBooNE error bars are statistical only. These particular signal distributions correspond to 20$\sigma$ (top panel) and 6$\sigma$ (bottom panel) discrepancies with the data.}
    \label{fig:diphotonDistribution}
\end{figure}

For the decay production channel, we generate a Monte Carlo (MC) sample of the scalars produced in $K/\pi$ decays by sampling the available phase space of the decays, and weighting each event by the squared matrix element. The amplitude and branching ratio formulas are given in Ref.~\cite{Krnjaic:2019rsv}, and our sampling method agrees with their reported $K \to \mu\nu S$ branching ratios. The pion MC is generated using the same amplitude, with the general substitution $K \to \pi$. The kinetic energy distribution of scalars produced via these modes, for $m_S = 24$ MeV, is shown in Fig.~\ref{fig:distribution}. The locations of the decays are shown in Fig.~\ref{fig:geant} (top and middle panels).

For the scalars produced by muons scattering in the beam dump, we use the \texttt{FeynRules}~\cite{Alloul:2013bka} Mathematica package to construct the Feynman rules for muon-philic interactions in \Eq{eq:lag}. This model is then used as input to the \texttt{CalcHEP}~\cite{Belyaev:2012qa} event generator to generate scalar emission $\mu N \to \mu N S$ cross sections (Fig.~\ref{fig:Feynman}, center) and MC events. These events are then re-weighted according to the electromagnetic form factor of the iron nucleus (see Ref.~\cite{Forbes:2022bvo} for a more detailed treatment). The probability of emitting a scalar in a muon-nucleus scattering event, as a function of muon energy incident on the beam dump, is
\be
    P_{\rm emit}(E_\mu) \simeq \frac{n_{N}}{\langle dE / dx \rangle} \int_0^{E_\mu} \sigma(E) dE , 
    \label{eq:ProbSprod}
\ee
where $n_N = 8.6 \times 10^{22}/{\rm cm}^3$ is the number density of nuclei in the iron beam dump, $\sigma(E)$ is the energy-dependent $\ \mu N \to \mu N S$ cross section, and $\langle dE/dx \rangle$ is the average energy loss of the muon in the medium, conservatively taken to be 16~MeV/cm \cite{ParticleDataGroup:2020ssz}. The typical path length in the beam dump is thus $l \sim 50$\,cm \cite{Groom:2001kq}. For each muon produced by the \texttt{GEANT4} simulation, the differential scalar emission probability in \Eq{eq:ProbSprod} is used to generate a random scattering event. A CalcHEP MC event of matching incident muon energy is randomly selected to generate the scalar 4-momentum. The production locations for scalars with $m_S = 100 \ {\rm MeV}$ are shown in Fig.~\ref{fig:geant} (bottom panel).

Example energy distributions of the mesons and scalars for $m_S = 24 \ {\rm MeV}$ are shown in Fig.~\ref{fig:distribution}. Using the 4-momenta of the scalars generated via the meson decay and scattering channels, we then evaluate the probability of $S$ decay within the fiducial volume of MiniBooNE/MicroBooNE, via the loop diagram in Fig.~\ref{fig:Feynman}, right. This is computed using the distance from the emission event to the detector, its line-of-sight width along the direction of the scalar momentum, and the decay length. The approximate decay length is given in Eq.~\ref{eq:approxD}, but in our analysis we use the full expression~(\ref{eq:width}) for each value of $m_S$. Diphoton invariant mass distributions for two representative scalar masses are shown alongside data and background expectations in Fig.~\ref{fig:diphotonDistribution}. We find that the largest signal rate given the MiniBooNE and MicroBooNE target and detector geometry comes from boosted $K^{\pm} \to \mu^\pm \nu_\mu S$ decays. Pion decays $\pi^{\pm} \to \mu^{\pm} \nu_\mu S$ are only kinematically allowed for $m_S < m_\pi - m_\mu = 34 \ {\rm MeV}$, but even at $m_S = 24 \ {\rm MeV}$ (Fig.~\ref{fig:diphotonDistribution}, bottom) the kaon decays still dominate. The contribution to the total $\gamma \gamma$ rate from muon scattering is largest at $m_S \sim 30 \ {\rm MeV}$, but is generally subdominant over the whole relevant mass range; unlike the meson decay channels, the scattering rate depends on the beam dump thickness and density, and thus this contribution may be more important in other experimental setups.

To evaluate our sensitivity to the $S$ decay diphoton signal, we use the fact that the same final state was used in MiniBooNE and MicroBooNE's measurements of neutrino-induced neutral current (NC) $\pi^0$ production, $\nu_\mu N \to 
\nu_\mu N \pi^0 \, (\pi^0 \to \gamma \gamma)$. In both analyses, the diphoton invariant mass $m_{\gamma \gamma}$ is reported, and we recast these analyses as searches for resonances with $m_{\gamma \gamma} = m_S$. Notably, the NC$\pi^0$ signal definition was slightly different for the two experiments. In MiniBooNE, the measurement included events with $\geq 0$ visible protons since MiniBooNE was insensitive to proton(s) at the interaction vertex~\cite{MiniBooNE:2009dxl}. MicroBooNE was able to separate their measurement into NC$\pi^0$ events with either 0 or 1 associated visible protons~\cite{MicroBooNE:2022zhr}. We consider the 0~proton case in MicroBooNE here since no protons are expected in an $S$ decay. 

The MiniBooNE (MicroBooNE) measurement was performed in neutrino  mode with $6.46 \times 10^{20}$ ($5.89 \times 10^{20}$)~POT and utilized a fiducual mass of 450~tons (77~tons). MiniBooNE's measurement in antineutrino mode is not considered here for simplicity. For MiniBooNE, we impose a ``track mass'' cut, which is used to separate diphoton and single-electron events, providing a natural lower bound on sensitivity, 
\be
m_{\gamma \gamma} \equiv \sqrt{2 E_1 E_2(1- \cos\theta_{12})} > f(E_1 + E_2),
\ee
where $E_{1,2}$ are the energies of the two photons, $\theta_{12}$ is their separation
angle in the lab frame, and the specific function $f(E_1 + E_2) \approx 32\,$MeV is given in Ref.~\cite{Karagiorgi:2010zz}.  For MicroBooNE, we employ the minimum photon energies used in their analysis ($E_{\rm high}>30$~MeV and $E_{\rm low}> 20$~MeV), but allow arbitrary separation angles. The smeared signal distribution of $m_{\gamma \gamma}$ is approximated as a Gaussian centered at the scalar mass with $15\%$ (33\%) resolution, consistent with the measured MiniBooNE (MicroBooNE) $\pi^0$ invariant mass in this region~\cite{Patterson:2009ki, MiniBooNE:2009dxl}. We assume that the 2$\gamma$ detection efficiency is a flat 40\% (6\%) across the invariant mass range used in this analysis, 32--128 MeV (10--120 MeV) in MiniBooNE (MicroBooNE), which is reasonably consistent with Refs.~\cite{MiniBooNE:2009dxl,MicroBooNE:2022zhr}. There is minimal sensitivity above these mass ranges primarily because the scalar decays before reaching the detectors. 

The lower threshold at MicroBooNE is noteworthy. Since MicroBooNE does not expect any true neutrino-induced $\pi^0$ (2$\gamma$, 0p) events in the 0--10 MeV bin~\cite{MicroBooNE:2022zhr} and it is difficult to estimate the detection efficiency behavior within this region, we have conservatively set the lower threshold to be 10 MeV. However, in principle, LArTPCs are sensitive to invariant masses below this value and we therefore encourage these experiments to explore sensitivity to $m_{\rm \gamma \gamma} < 10 \, {\rm MeV}$, where pion decay could dominate and for which there is a large swath of allowed parameter space that could be used to explain the $g-2$ anomaly.

The primary background for our analysis is $\pi^0$ production in neutrino-induced NC$\pi^0$ scattering, and in setting our limits we compare our predicted signal yield against the observed diphoton distributions from Refs. \cite{MiniBooNE:2009dxl,MicroBooNE:2022zhr}, noting that no significant excess is apparent in the datasets. For MiniBooNE, we calculate the significance of an injected signal ($s_i$) with a standard $\chi^2$ test, 
\be
    \chi^2 = \sum_{i,j={\rm bins} } (d_i-s_i-b_i)M_{ij}^{-1}(d_j-s_j-b_j)~,
\ee
where $d_i$ is the observed data, $b_i$ is the background prediction, and $M_{ij}$ is the uncertainty matrix. We construct $M_{ij}$ using the statistical uncertainties on the data and assume fully correlated bin-to-bin systematics associated with flux (12.4\%), cross section (8.4\%), and the detector (5.1\%), consistent with Ref.~\cite{MiniBooNE:2009dxl}. We note that these systematics are nominally applicable to MiniBooNE's NC$\pi^0$ analysis, rather than a generic 2$\gamma$ search, and that the detailed correlations between bins may be important. For MicroBooNE, we use a Poisson extended maximum likelihood 
 $\chi^2$ definition~\cite{ParticleDataGroup:2020ssz,cowan1998statistical} since the number of reported events is $<20$ across many of the relevant bins:
\be
\chi^2_{\rm poiss} = 2 \sum_{i= \rm bins}\Big[N_{ \text{pred},i}-d_i+d_i \ln(d_i/N_{ \text{pred},i}) \Big],
\ee
 where $N_{ \text{pred},i}=s_i+b_i$. This MicroBooNE estimate is a ``statistical uncertainty only" approximation, which ignores systematic uncertainties. However, the relevant reported invariant mass bins (10--120 MeV) are dominated by statistical errors and a more complete treatment is therefore not expected to markedly change our reported limit. A more detailed study by the collaboration, in particular to search for generic 2$\gamma$ (rather than NC$\pi^0$) events, validate the $\chi^2$ thresholds with fake data studies (given the low statistics), and include systematics, would be valuable. We define a $5\sigma$ exclusion limit in the region where the $p$-value associated with the $\chi^2$ falls below $2.8 \times 10^{-7}$.

\label{sec:search}

\section{Results}
\begin{figure}[t!]
\hspace{-0.5cm}
    \includegraphics[width=3.5 in]{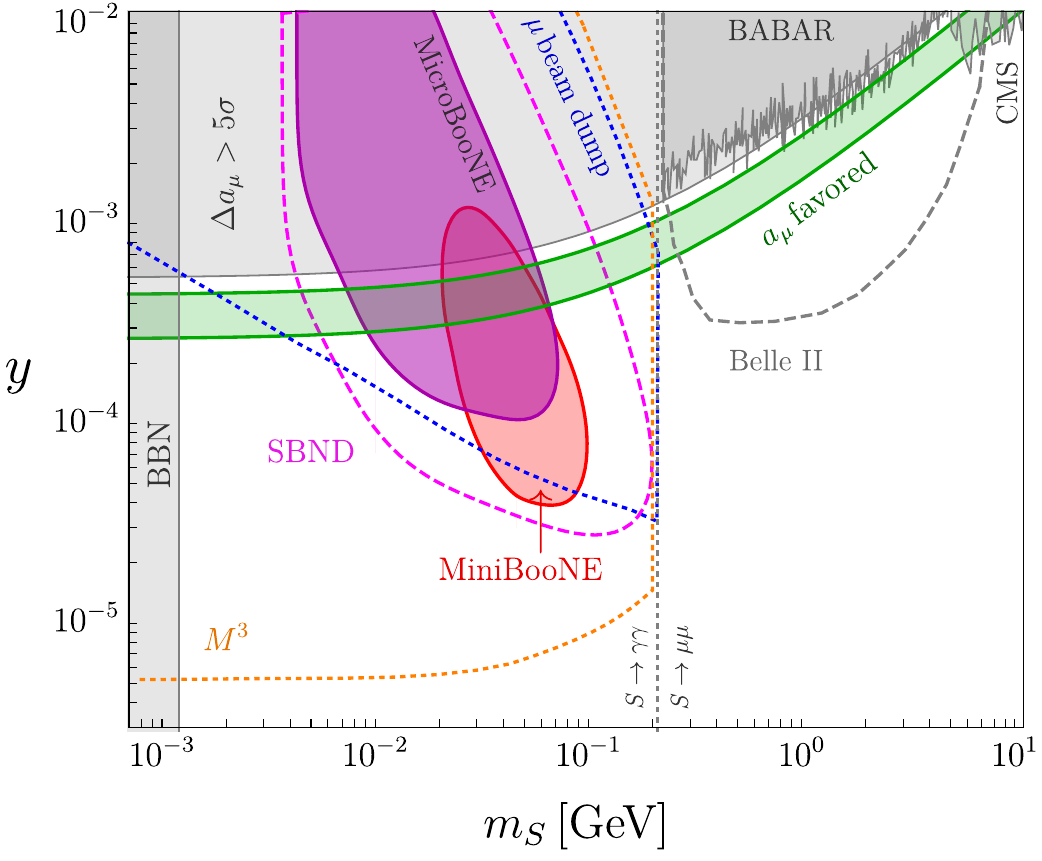}~
    \caption{Parameter space for 
    the singlet scalar in \Eq{eq:lag}, decaying to diphotons via \Eq{eq:width}. The red and purple shaded regions are excluded at $5\sigma$ by existing MiniBooNE and MicroBooNE data, respectively, based on the analysis in this paper and the magenta dashed curve represents a future $2\sigma$ projection for  SBND sensitivity. The green band represents the parameter space where this scenario resolves the muon $g-2$ anomaly. For $S > 2m_\mu$ we show constraints from BaBar and 
    CMS based on the $S \to \mu^+\mu^-$  decay channel along with future projections for Belle-II \cite{Capdevilla:2021kcf}. For $S < 2m_\mu$ we also show projections for a dedicated muon beam-dump experiment \cite{Chen:2017awl} and for the $M^3$ muon missing momentum proposal \cite{Kahn:2018cqs}. Also shown are cosmological bounds for $m_S < 1$~MeV where 
 $S$ particles thermalize with the SM as a relativistic species during Big Bang nucleosynthesis \cite{Krnjaic:2019dzc}.
    }
    \label{fig:results}
\end{figure}

In Fig.~\ref{fig:results} we show our 
sensitivity in the coupling $y$ versus mass $m_S$ parameter
space. Our main results are the red shaded region which represents the MiniBooNE exclusion at $5\sigma$ confidence, and the purple shaded region with the analogous interpretation for MicroBooNE. We also include a projection for the SBND LArTPC-based experiment at Fermilab~\cite{MicroBooNE:2015bmn}, set to take first data in 2024~\footnote{This projection is not to be considered as coming from the SBND collaboration.}. We consider a realistic detector geometry (77~ton fiducial mass) at a distance of 60~m from the beam dump, $10^{21}$~POT, and use the same invariant mass resolution and efficiency assumptions as MicroBooNE. The excluded regions all have similar characteristic shapes. At large coupling and high mass, the decay length is much shorter than the detector baseline, and the signal flux is exponentially suppressed. In the low-coupling region, scalar production becomes increasingly unlikely. In the low-mass limit, the analyses are no longer suited to detect the diphoton signal. It is noticeable that SBND has a much shorter baseline - this allows it to exclude further into the shorter lifetime region. It also has a higher geometric acceptance, which allows it to extend further into the small coupling regime. The SBND NC$\pi^0$ background estimate comes from scaling MicroBooNE to the SBND location using the ray tracing simulation, which estimates an increase in background event rates of a factor of $N_{\rm bkg, SBND} \approx 90 N_{\rm bkg, MicroB}$. This reasonably agrees with the simple estimate $N_{\rm bkg} \propto N_{\rm POT} V / r^2$, with $V$ the detector volume and $r$ the baseline. 

The green band in Fig.~\ref{fig:results} represents the favored region for reconciling the anomalous observed experimental value of $g-2$ to within $2\sigma$ agreement \cite{Muong-2:2023cdq}. The combination of MiniBooNE and MicroBooNE constraints exclude scalars of mass between 10 and 60 MeV in this band. The gray shaded region directly above the green band is excluded by muon $g-2$ measurements, as loop contributions from $S$ would contribute beyond the observed value and reintroduce a discrepancy between theory and experiment at the $>5 \sigma$ level. For $m_S > 2m_\mu$, Fig. \ref{fig:results} also shows constraints from $S \to \mu^+\mu^-$ decays assuming a $100 \%$ branching fraction into this channel. 
We show limits from the BaBar $e^+e^- \to 4\mu$ search dashed projections for improving $B$-factory limits with future Belle-II analyses \cite{Capdevilla:2021kcf}.  For $m_S >$ few GeV, there is also a CMS constraint based on reinterpreting a $Z \to 4\mu$ decay search \cite{Capdevilla:2021kcf}. 

Below the dimuon mass threshold, Fig.~\ref{fig:results} also shows several projections for future searches. The blue dashed region is the projection for a dedicated muon beam-dump experiment proposed in
Ref.~\cite{Chen:2017awl}, which could cover much of the remaining $g-2$ band. The orange dashed curve is based on a proposed muon missing momentum ($M^3$) search strategy \cite{Kahn:2018cqs}. Although $M^3$ is nominally sensitive to invisibly-decaying particles produced in muon interactions, the scalar $S$ can be sufficiently long-lived on the length scales of the proposed experiment such that it is effectively invisible, and $M^3$ would probe the same parameter space as the model considered here. As mentioned earlier, access to the $0-10 \ {\rm MeV}$ bin in $m_{\gamma \gamma}$ at MicroBooNE/SBND would allow us to place strong limits on $m_S < 10 \ {\rm MeV}$ from $S$ production in charged pion decays, which would likely be competitive with the dedicated muon beam dump projections.

\section{Conclusions and Outlook}
The results of this work should be taken as a proof-of-concept validation of the potential of reinterpreting data collected at neutrino short-baseline experiments in the context of muon interactions beyond the SM. 
There are some experimental subtleties that were not taken into account in this analysis, and thus we expect the results to be correct to within an $\mathcal{O}$(1) factor.
Nonetheless, given the amount of parameter space which is accessible given current data, espeically the highly-motivated $g-2$ region, we strongly urge the MiniBooNE and MicroBooNE collaborations to undertake a generic 2$\gamma$ search (including systematics and detailed bin-to-bin correlations not present in publicly-available data) in order to achieve more accurate results.
Futhermore, we encourage other neutrino experiments producing copious auxiliary muons---such as NOvA~\cite{osti_935497}, T2K~\cite{T2K:2011qtm}, DUNE \cite{DUNE:2015lol}, and ESS~\cite{ESSnuSB:2013dql}---to undertake similar muon studies to maximize the physics potential of the experimental infrastructure. 

More generally, in this {\it Letter} we have illustrated the potential of neutrino experiments to also function as probes of weakly-coupled new physics produced via muon interactions.
The production of muons associated with neutrinos is completely generic with a primary proton beam, and thus the secondary muon source is a byproduct of the neutrino production and is generated at no additional cost to the experiment. 
Not only is a muon program at neutrino sources economical, but it is also uniquely sensitive to muon-philic new physics scenarios, which is an under-explored region of parameter space due to the lack of dedicated muon sources.

Finally, we emphasize the practical importance of these secondary muon programs: the costs are small and the required timescales are short. 
An investment in exploring the implications from muon physics at these experiments can not only inform and shape the future construction of other large-scale experiments, but also could provide the first hints towards understanding the persistent puzzles of the SM.

\section*{Acknowledgments}
We would first like to thank the organizers and participants of the 2023 ACE Science Workshop at Fermilab, where the idea for this work was conceived, for stimulating discussions and feedback. C.C. is supported by the U.S. Department of Energy (DOE) Office of High Energy Physics under Grant Contract No. DE-SC0012567. The work of Y.K. was supported in part by DOE grant DE-SC0015655. J.S. is supported by the Department of Energy, Office of Science, under Award No. DE-SC0007859.

\bibliographystyle{utphys3}
\bibliography{biblio}

\end{document}